\documentstyle[11pt,newpasp,twoside,epsf]{article}
\def\emphasize#1{{\sl#1\/}}
\def\arg#1{{\it#1\/}}
\let\prog=\arg

\def\edcomment#1{\iffalse\marginpar{\raggedright\sl#1\/}\else\relax\fi}
\reversemarginpar

\begin{document}
\title{Recent Observations of X-ray Emission from Galactic Binary Neutron Stars}
 \author{Jean Swank}
\affil{NASA/Goddard Space Flight Center, Code 662, Greenbelt, MD 20771}

\begin{abstract}Recent measurements of young accreting binary
neutron stars are determining more 
precise magnetic field and accretion parameters. 
A low magnetic field accreting, millisecond 
pulsar has finally been found in a binary burster.  
At least 20 low-mass binaries have exhibited high frequency
oscillations, 300--1200 Hertz. The majority
have, for some range of luminosity, the pair of quasiperiodic
oscillations that have never been seen in a black hole candidate.
Recent evidence from burst oscillations 
strengthens the case that the difference frequency of
this pair is close to the spin frequency of the neutron star. These
oscillations  are correlated
with the spectra, luminosity, and low frequency oscillations. 
Quasiperiodic oscillations are also seen sometimes in strong
magnetic field pulsars, where their origin can be closely examined.

\end{abstract}

\section{Introduction}

It is now believed that we have galactic binary neutron stars with
magnetic fields that are as low as $10^8$ Gauss and as high as $5
\times 10^{14}$ Gauss. There are single neutron stars with magnetic
fields as low, and probably single neutron stars with a field as
high. Because of recent measurements it is now possible to plot the
magnetic fields and rotation periods of representative members of the
varied populations. These are displayed in Figure 1, along with the
fields and periods of some related objects. The evolutions of systems
from birth of the neutron stars to their place in this diagram are
subjects in themselves. Only now are the X-ray measurements
establishing the places of the objects with reasonable accuracy rather
than as an order of magnitude hypothesis.

There are many aspects of the accretion onto the binary neutron stars
that are not yet understood. I will touch upon some of these. But
observations by RXTE and BeppoSAX have established some facts that
vindicate former years of assumptions, enable calculations to be
done that depend on the parameters that were determined, and lead to a
new round of questions.

\section{Pulsars}

\subsection{Cyclotron Resonance Features in High Magnetic Field Pulsars}
The "classic" accreting binary pulsars are those that were first found
in X-ray astronomy, with rocket flights and UHURU. Her X-1 with its
1.24 s period and Cen X-3 with its 4.85 s period appeared to exhibit
long term speed up at about the rate that the observed accretion would
imply, if the magnetic field at the surface were about $10^{12}$
Gauss, and this was close to the fields of young pulsars like the Crab
pulsar, as obtained from their spin down rates. The first detection
of a spectral feature that could be associated with transitions in a
strong magnetic field was twenty years ago (Tr\"{u}mper et al. 1978).  The
Her X-1 flux was falling fast with energy at the energy of the feature
and it was uncertain whether absorption or emission was being seen. By
now the significance of the feature and the ability to measure it at
different phases of the pulse period have established that it is
usually in absorption. A few other sources exhibited cyclotron
resonance features, in particular during a transient outburst of
3U~0115+63 the first evidence of a fundamental and first harmonic
appeared. Ginga 
detected a feature in enough sources to suggest a connection between
the continuum spectrum and the cyclotron resonance feature 
(Mihara, Makishima, \& Nagase 1998). In 1996,
Beppo SAX and RXTE both fielded instruments with special sensitivity
to cyclotron lines in the energy range 10-100 keV.

\begin{figure}
\plotfiddle{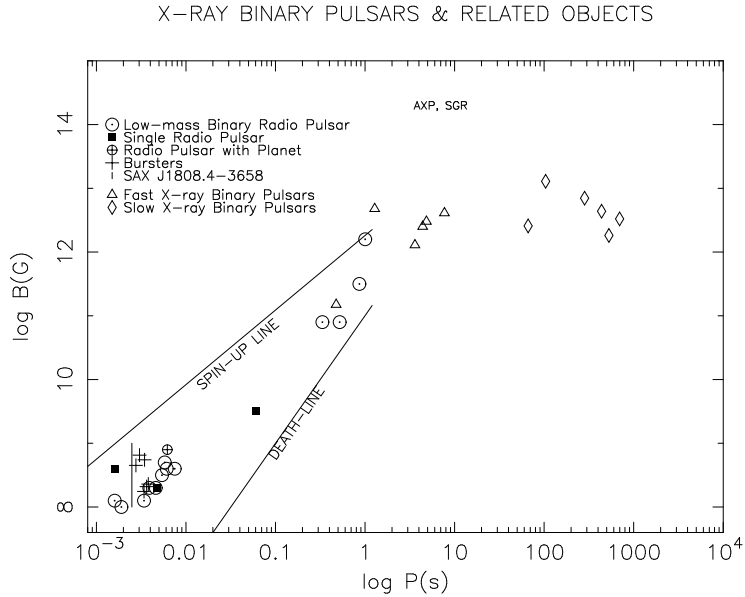}{3in}{-90}{40}{40}{-150}{230}
\caption{X-Ray Binary Pulsars and Related Objects. Accreting pulsars
are shown for which magnetic fields are estimated, from
cyclotron features for periods longer than 1 s and on the basis of
assumed equilibrium for periods shorter than 10 ms (White \& Zhang 1997). 
Radio pulsars
expected to have come from low-mass binary X-ray pulsars
are shown. The tentative B-P region of ``Anomalous
X-ray Pulsars''(AXP), and the ``Soft Gamma Repeaters'' (SGR) is labeled.}

\end{figure}

The number of detections has risen and some systematic correlations
are emerging which are beginning to test theories of the emission
region. The results for the cyclotron absorption energies and the 
assumed dipole magnetic field are summarized in the table. 

\begin{table}
\caption{Cyclotron Lines and Magnetic Fields}
\begin{tabular}{llrrrrrll}
\tableline
P(s)  &Source$^1$	&E1$^2$	&E2	&E3	&E4  &B$^3$ &Mission$^4$&\\
\tableline
1.28  &Her X--1	&42.1   &       &       &    &4.8   &Balloon,H,G,X,S&\\
3.61  &(T)4U 0115+63 & 12.4 &21.4 &33.6 &48.5 & 1.1 &H,G,X,S&\\
4.4 &(T) V 0331+53 & 27.2 & & & & 2.5 & G&\\
4.85& Cen X--3&27.9&&&&3&S&\\
7.7& 4U1626-67&37&&&&4.1&S,X&\\ 
66& (T)Cep X--4&28.8&&&&1.2&G&\\
104& (T)A 0535+262&55&115&&&13&C&\\
283& Vela X--1&24.5&54.4&&&7.0&G, S, X&\\
438& 4U 1907+09&18.9&39.4&&&4.4&G, S&\\
529& 4U 1538--52&20.6&&&&1.8&G&\\
696& GX 301--2&37.6&&&&3.3&G&\\
\tableline
\tableline
\end{tabular}\\
{\footnotesize $^1$Names of transient sources are preceded by (T).  
$^2$Energy in keV at centroid of feature. 
$^3$Surface dipole field in $10^{12}$ Gauss.
$^4$Missions carrying instruments which showed evidence 
for one or more cyclotron features in the spectra were, 
after the balloon flights, HEAO--1 (H), Ginga (G), CGRO (O), RXTE (X), 
and BeppoSAX (S). See Mihara et al. 1998, 
Dal Fiume et al. 1999, Heindl et al. 1999, and Santangelo et al. 1999.} 
\end{table}

Three of the sources in the table are transients and the absorption
features, especially the higher harmonics, are observed when the
source is bright, presumably because the increased accretion flow is
necessary for sufficient optical depth in the absorption feature.

The observations of harmonics at about the correct spacing confirms
expectations about the energy levels of electrons in the strong
magnetic fields. While the agreement is to 10 -- 15 \% the best fits
do not confirm an exact integral ratio. However the ratio is not expected to
be exactly integral and there are reasons for different harmonics to
be formed at different heights. It seems plausible that the
complexity of the accretion process could easily lead to sufficiently
different locations for particular processes to dominate to explain
the differences. When the data are good
enough, the fits to spectral models require a high energy cutoff in
addition to the resonance absorption.

The widths of the observed features are tens of keV ($E/ \Delta
E \approx$ 2 -- 4).  The full-width-half-maxima appear to be
proportional to the energy of the fundamental and consistent with
Doppler broadening in thermal plasma with kT = 3 -- 20 keV (Dal Fiume et
al. 1999).  The observations that have already taken place contain
more detailed information about the behavior of the features as a
function of spin phase and should soon significantly constrain
detailed pictures of the flow of accreting gas onto the neutron star. 

\subsection{Quasiperiodic Oscillations}

Under some circumstances there are  strong quasiperiodic oscillations (QPO)
in the flux from accreting pulsars (Kommers
\& Chakrabarty 1998; Heindl et al. 1999; Finger 1998). 
Two reports of
detailed studies of QPO in RXTE data have been inconsistent with the
magnetic beat precession model which was first thought likely to be the
explanation. In these cases the cyclotron line now provides
independent knowledge of what the magnetic field is. With this
knowledge, it appears that there is structure in the disk outside the
magnetosphere, in structures in the disk.  During the recent outburst
of 3U~0115+63, quasiperiodic oscillations appeared at 2 mHz with a 5
\% rms amplitude. The 48 mHz QPO in 4U~1626--67 
also suggests blobs in Kepler motion outside the Alv\`{e}n shell.

While the very low frequency oscillations suggest origins outside the
magnetosphere, evidence for characteristic time-scales of the polar cap
or the magnetic field near the surface of the neutron star are sought
at frequencies of 10--1000 Hz. For an observation for a binary orbit
of Cen~X--3 when the source was very bright, Jernigan, Klein, \&
Arons (2000) found a broad peak in summed power spectra which they
find comparable to results of simulations of accretion onto a
magnetic pole in which ``photon bubble oscillations'' are excited.

\subsection{Weak Magnetic Field Pulsars} 

More than fifty X-ray sources, mostly in the galactic bulge, have
faint optical counterparts and are presumed to be low-mass X-ray
binaries which have been accreting for the lifetime of the low mass
companion,  some $10^8$ yr. Since pulsations had not been found
in deep observations by HEAO-1, EXOSAT, Ginga, and other missions even
when time resolution  could have detected periods as short as a
few milliseconds, it has been clear that the neutron stars and
accretion flow were different from those in the young, high-mass X-ray
binaries. The weak fields of older radio pulsars have suggested that
the old neutron stars in X-ray binaries have correspondingly low
magnetic fields. Exactly how weak they are is still uncertain, but
probably not weak enough to completely explain why the X-ray flux
produced in accretion is not pulsed. Sufficient optical depth to
scattering may be able to explain it, but in that case lower
luminosity sources with lower accretion rates may have less obscured
lighthouse beams.

Now we have one bona fide pulsing burster, XTE~J1808--3658. In June 1998,
RXTE's PCA field of view serendipitously scanned over a transient
in the galactic bulge (Marshall 1998). It appears to be the same source as
SAX~J1808.4--3658, discovered in an outburst
18 mo. before (in't Zand et al. 1998). Follow-up observations 
discovered (Wijands and van der Klis 1998) 
that it had coherent pulses at a frequency of 401 Hz. The observed
period varied, as Chakrabarty and Morgan proved (1998), with the
Doppler shifts of a 2 hour binary. During the transient's first appearance,
two
bursts occurred, with fast rise  and an
exponential cooling decay that is the definition of Type I bursts
that are believed to be thermonuclear flashes on a neutron star. This
is in contrast to the ``Bursting Pulsar'' GRO J1744-28, which has a
long pulse period (0.5 s) and bursts that fit the mold of Type II
bursts thought to be from an accretion instability.

Extensive PCA observations of the source traced its 25 day decay. At a
distance argued to be 4 kpc (from the peak luminosity of the bursts),
it had a peak luminosity of $5 \times 10^{36}$ ergs s$^{-1}$ and was
still pulsing in the last PCA observations at $4 \times 10^{34}$ ergs
s$^{-1}$. Neither the spectrum nor the pulsed fraction appeared to
change. From these facts were deduced a field
of $5 \times 10^8$ Gauss, along with  constraints on the mass
radius relation (Psaltis \& Chakrabarty 1998; Burderi \& King 1998)

Chakrabarty \& Morgan found a very small mass
function. Although the signal is very strong, the value of $a \sin(i)$ (
62.8 lt-ms) would suggest an improbably small inclination angle ($i$) unless
the companion is less massive than  a main sequence star
filling its Roche lobe in a 2 hour orbit. If it turns out that we
can only see the pulsations looking at the system face on,
we may never find very many.

However it is possible that the fields of brighter sources are weaker
or the poles obscured and that SAX~J1808.4--3658 is the first to be
found of a population of sources. It is clear from BeppoSAX Wide Field
Camera results that there are in the galactic bulge many transient
bursters, which as transient persistent sources do not get brighter
than about 100 mCrab. Some of those may also be fast pulsars. In view
of that possibility, we have been making biweekly scans of the
galactic center region. These PCA scans have detected new transients
down to a level of a few milliCrab, some  with very interesting
variability (e.g. XTE~J1819--25=V4631~Sgr); but so far no coherent 
pulsations have
been found.

\section{Non-Pulsing Low Mass X-ray Binaries} 

The 50 low-mass X-ray binaries divide into the luminous Z sources
(only 6 sources), the luminous ``atoll'' sources ( 5 agreed
upon), the less luminous and regularly bursting atoll sources (at
least a dozen), and over 20 sources which are bursters but have weak
persistent flux. Z and atoll sources are both capable of bursting, in
that GX 3+1 among the atolls, and Cyg X-2 among the Z sources do
sometimes burn unstably and exhibit bursts.

The kilohertz pulsations that RXTE discovered have appeared in persistent
flux measurements in all the Z sources, in none of the
bright atoll sources, in all the atoll bursters, and in some of the
faint bursters.
Bursts have been observed from many bursters. So far, oscillations
have been seen in bursts from 6 of them. The kilohertz frequencies seen
in the persistent flux are not seen in the burst flux. Vice-versa, the
burst oscillations have not been seen in the persistent flux.

The QPO seen in the persistent flux from these sources change in
frequency by large amounts (factor of 2) and none of the models assign
the rotation of the neutron star to these QPO periods. The QPO during the
bursts change by less than about 1 \% and in the only detailed model
the period is the rotation of the neutron star. These oscillations
strongly suggest that  even the sources which do not have coherent high
frequency pulsations are fast rotating neutron stars.

\subsection{Burst Oscillations}

The sources from which bursts have exhibited oscillations remain
4U~1728--34 (363 Hz), 4U~1636--53 (581 Hz), 4U~1702--43 (330 Hz), 
Aql~X--1 (549 Hz), KS~1731-26 (524 Hz) , and
MXB~1743--29 (589 Hz). Not all bursts from these exhibit oscillations. 
Figure 2 shows
spliced data segments containing 6 bursts, in which 5 of the 6 exhibit
the same asymptotic frequency and similar variations by a few Hz in
the frequency as the burst progresses. It is not understood why they
are sometimes not present. It remains plausible that the burst sources
for which no burst oscillations have been found, still are
fast rotating neutron stars, and may at some time exhibit
oscillations. EXOSAT observations of bursters established that these
sources have cycles of different burst characteristics.

\begin{figure} 
\plotfiddle{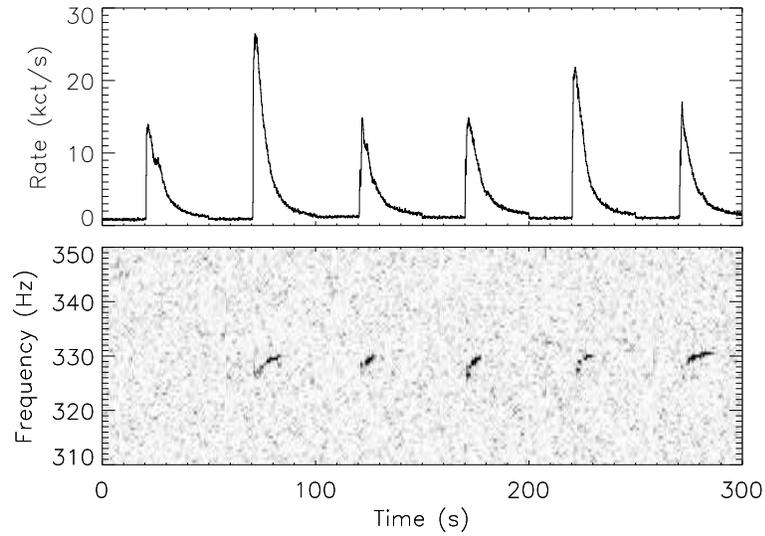}{3in}{0}{55}{55}{-160}{-160}
\caption{Burst Oscillations in An Atoll Burster (C. Markwardt). 
Sections of data containing
bursts from 4U 1702-43 are adjoined in the top panel. 
In the bottom is a grey scale dynamical power spectrum 
showing the frequency range of the burst oscillations.  
}

\end{figure}

\begin{figure}
\plotfiddle{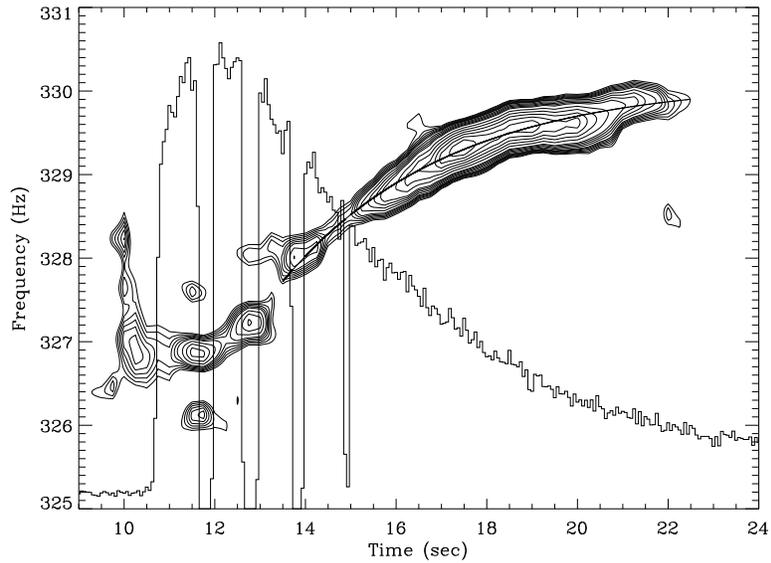}{3in}{90}{45}{45}{180}{-30}
\caption{Variable Frequency of Oscillations during Burst (T. Strohmayer). 
The burst light curve is superposed. (Gaps occur where 
the data readout was not fast enough for this high time resolution data.)}
\end{figure}

There is evidence that despite the small, but definite change in
frequency in a burst, that at least the oscillations during the
cooling tails of bursts are coherent.
Strohmayer and Markwardt (1999) showed that
for a simple  frequency function of time $\nu (t)$,  the Fourier 
power is maximized with the
phase of photons arriving at time $t_{\jmath}$ given by:
$\phi _{\jmath} = 2 \pi \int _{0}^{t_{\jmath}} \nu (t')dt'.$
Figure 3  shows contours of smoothed 
power in 2 s intervals for the second  burst in Figure 2. 
The frequency  function of time shown by the line 
is a good fit to the assumption of coherence during that interval.

In a number of bursts the time dependence of the frequency suggested
by the power spectra is more complex. A very strong signal during the
initial rise of the burst can be interpreted in terms of 1 or 2 hot
spots growing in size as the burning area increases to cover the whole
star (Strohmayer, Zhang, \& Swank 1997) . There is often a gap between
that signal and the signal appearing in the burst tail. An asymmetry
on the neutron star could lead to a single hot spot. If the star has
an approximately dipole field, two poles might be expected to stand
out as hotter than the surroundings and the observed frequency could
be the harmonic of the spin frequency. Miller (1999) found that for
4U~1636--53 there was evidence for the subharmonic (290 Hz) of the
dominant signal, briefly at the beginning of some bursts. Otherwise,
subharmonics and harmonics have amplitudes more than an
order of magnitude less than the main frequencies.

\subsection{Persistent Flux QPO}

The typical kilohertz QPO signals in the persistent flux are 2 peaks, one
narrower than the other, which move in frequency as the luminosity
moves, with time scales of hours. One or both of the features may
become broad and faint for a time. When the frequency wanders and both
are present, they wander together. They keep approximately the same
interval between them, although this interval appears to shrink by as
much as 20 \% at the upper end of the  ranges that have been observed.

Both the Z sources, the bright atoll sources and the atoll burst
sources have substantial structure in the lower frequency part of
power spectra, below 100~Hz. This is where the QPO discovered with
EXOSAT appeared, with their correlation with the luminosity.  For each
Z source (See Wijnands and van der Klis 1997) when the source is on its
``Horizontal Branch'', both Horizontal Branch Oscillations (HBO)
and kilohertz QPO  are present and
their changes with luminosity are correlated. The HBO is different
from the kilohertz QPOs in having a strong harmonic, while there has
been no evidence for any harmonics or subharmonics for the kilohertz
oscillations. It is characteristic of the power spectra on the
horizontal branch of Z sources to have band limited white noise at 
frequencies below a break frequency, with the HBO  just above the
break. All of these features are usually correlated with the
luminosity on that branch.

Qualitatively the atoll bursters have not only kilohertz oscillations
similar to the more luminous Z sources, but also similar low frequency
power spectra. They have band limited white noise, which was a defining
characteristic of atoll sources  and above
the break they have a strong additional QPO feature. Their HBO are
broader, with less clear harmonic structure and more variation in the
low frequency QPO. Ford and van der Klis (1998)document
that for 4U~1728--34 the break frequency and what they call the ``lower
frequency Lorentzian'' are strongly correlated. 
The correlation $\nu _L \propto \nu _u ^2$ is just the 
relation that Lense-Thirring precession would imply for 
clumps in Kepler motion in tilted circular orbits  close 
to the innermost circular orbit (Stella \& Vietri 1998).

Consequences of models of the kilohertz QPO have been under
development as the observed characteristics are being explored.
These very different models each involve some physics which appears
relevant. Each also has some points of possible difficulty.

The sonic point model (Miller, Lamb, \& Psaltis 1998) provides a mechanism for
the Kepler frequency and the rotation frequency of the neutron star to
beat, without assuming a strong magnetic field and without the
dependence on luminosity that the magnetic beat frequency model
implied. The sonic point determines the inner radius of the disk,
where blobs circulate at the Kepler frequency. The spin beats with the
Kepler blobs to produce the lower kilohertz frequency. But because the
HBO are also observed and appear to be well described by the magnetic
beat frequency model, it is currently supposed that the magnetosphere
balances accretion only partially. The model assumes that the
accreting neutron star does have poles that are hotter than the rest
of the star, but the effect of only one of these is observed in the
beats. The difference frequency between the two kilohertz frequencies
is approximately the spin frequency. It can be reduced however because
the radiation interacts with blobs that are spiraling into the
neutron star. 

Stella \& Vietri (1999) also pointed out that 
for slightly elliptical orbits, the epicyclic motion would
have a frequency which, if it  beats with the Kepler motion, could produce
the lower kilohertz frequency. While plausible eccentricity and neutron
star mass could fit the data for the best measured case (Sco X-1), the
magnitude of the Lense-Thirring precession would require neutron star
moment of inertia 2-4 times larger than for any derived equation of
state. No mechanism has been detailed for radiation
variation at these frequencies.

The sometimes very narrow QPO have also suggested a resonant
oscillation of some sort. In one model involving resonances,
(See Titarchuk, Osherovich, \& Kuznetsov 1999.)
the plasma in the boundary between
the disk and magnetosphere undergoes radial and vertical oscillations
as it circulates with the magnetosphere. In this case the lower
kilohertz frequency is identified with the Kepler frequency, the
higher kilohertz frequency with a plasma frequency called ``upper
hybrid'', and the HBO frequency with the low branch. Assumption that
the inner disk interacts with the boundary of the neutron star rather
than the magnetosphere can possibly explain  other
features in the power spectra.

\subsection{The Role of Accretion in the QPO}

A remarkable aspect of the kilohertz oscillations has been the fact
that the phenomenon and the numerical range of frequencies are
independent of the average luminosity of the sources. Thus the
frequencies range from 400 Hz to 1200 Hz for an atoll burster with
peak luminosity of perhaps $10^{37}$ ergs s$^{-1}$, just as it does for a
Z source with peak luminosity above $10^{38}$ ergs s$^{-1}$.  This
first suggested that the frequency must depend on something
independent of the accretion rate, such as the inner-most stable orbit
or the radius of the neutron star (Zhang, Swank, \& Strohmayer 1997).
However, for a given source, over
short ranges of luminosity for periods of time less than or on the
order of a day, the frequency increases with luminosity. 
Assuming that one of the frequencies is a Kepler
frequency, it should saturate at the frequency of motion in the
inner-most stable orbit. Observations of 4U~1820--30 have 
seemed to show this effect. (See recent confirmation by Kaaret et al. 1999).

Another aspect of this independence of luminosity between sources, 
is that a given source can go through the same frequency run for different 
ranges of luminosity. Mendez (1999) has put
together the data from 4 sources that show this characteristic.
On the other hand, the frequency varies systematically 
around the atoll in a color - color diagram. 
These characteristics suggest that for the atoll bursters there is a
well-defined emission geometry and mechanisms which produce the
QPO. The frequencies and spectra may be related to some mass flow rate
in the system. But the luminosity is
not uniquely related.

\subsection{Phase Lags}

The PCA has been used to measure the phase or time lags between
different energy bands for many sources and for many of the temporal
phenomena, the continuum noise in power spectra, various QPO in the
persistent flux and the oscillations in the bursts. Hard lags have
been known since EXOSAT and Ginga, to
characterize black hole candidate noise and HBO. They also appear to
characterize the LMXB noise (Ford \& van der Klis 1999). However the lower
of the two kilohertz frequencies is definitely subject to soft lags, 
although they are very small in comparison to
the soft lags of SAX~J1808.4--3658. Lags for the bursts and the upper
kilohertz QPO are small and the situation is not yet clear. 
While there are
possibilities of either Compton up or down scattering, and Doppler
shifts can cause lags, definitive patterns and consistent
interpretations are uncertain.

\section{Conclusions}

The new results on accreting binary neutron stars found by RXTE and
BeppoSAX provide new tools to gain physical insight both  into the compact
objects that are the last equilibrium point before collapse to a black
hole and into the binary accretion flows onto them.

In the case of the strong field accreting pulsars, cyclotron features
allow direct measurement of their magnetic fields and the timing
signatures of the flow then illuminate apparent irregularities 
in the disks.    The
conditions of their occurrence are not clear and they are far from
being well studied. There are hints of vibration signatures that might
come from the accretion mound above the polar cap and  are possible 
evidence for photon bubbles, but they need confirmation and more study.

Magnetic fields down near $10^8$ Gauss in low-mass X-ray binaries are
confirmed by the discovery of one irrefutable example.
The uncertainty of important parameters of the system leave it still
ambiguous whether we should find more such systems.  A burster
itself, it could resolve many technical questions about the QPO in
LMXB if it ever manifests them.

Evidence continues to support the conclusion that bursters are fast rotating
neutron stars with low fields. The burst frequency and the difference between
two high frequencies in the persistent flux seem rather clearly related. But the phenomena are still complex in detail. 

The persistent flux QPO around 1 kHz and around 100 Hz are correlated
and correlations appear very general, holding for atoll sources, Z
sources, and possibly black hole candidates as well. 
Competitive models are still working out
detailed implications for comparison. It is notable that they all
involve effects within a few kilometers of the neutron star, where
effects of General Relativity are
involved.

\end{document}